\documentstyle[aps]{revtex}
\begin{document}
\draft
\newcommand{\ket}[1]{|#1\rangle}
\newcommand{\bra}[1]{\langle#1|}
\newcommand{\scpr}[2]{\langle#1|#2\rangle}
%\wideabs{
\title{Trions in a periodic potential}
\author{Wojciech Florek}
\address{A. Mickiewicz University, Institute of Physics\\
ul. Umultowska 85, 61--614 Pozna\'n, Poland}
\date{May 7, 2000}
\maketitle
\begin{abstract}
 The group-theoretical classification of trion states 
 (charged excitons $X^\pm$) is presented. It is based on considerations
 of products of irreducible projective representations of the 2D translation
 group. For a given BvK period $N$ degeneracy of obtained 
 states is $N^2$. Trions $X^\pm$ consist of two identical particles (holes 
 or electrons), so the symmetrization of states with 
 respect to particles transposition is considered. There are $N(N+1)/2$ 
 symmetric and $N(N-1)$ antisymmetric states. Completely antisymmetric
 states can be constructed by introducing antisymmetric and symmetric
 spin functions, respectively. Two symmetry adapted bases are considered: 
 the first is obtained from a direct conjugation of three representations, 
 whereas in the second approach the states of a electrically neutral pair
 hole-electron are determined in the first step. The third possibility, a 
 conjugation of representations corresponding to identical particles in the
 first step, is postponed for the further investigations. 
\end{abstract}
%}
% body of paper here

\section{Introduction}

The quantum Hall effect and high temperature superconductivity have 
raised interest in properties of the two-dimensional electron gas subjected to 
electric and
magnetic fields. The observation of (negatively) charged excitons 
\cite{shields} has recalled a forty-year old concept of excitons ``trions''
or ``charged excitons'' introduced by Lampert in 1958 \cite{lam}.
Recently, such excitons, consisting of two holes and an electron or two 
electrons and hole (denoted $X^\pm$, respectively), have been investigated 
both experimentally and theoretically \cite{exper,theor,Dzy}. 

In this paper classification based on translational symmetry in the presence 
of a periodic potential and an external magnetic field is presented. To 
perform this task the so-called magnetic translation operators, introduced by 
Brown \cite{brow} and Zak \cite{zak}, are used. These operators commute with 
the standard Hamiltonian of an electron in the magnetic field 
${\mathbf H}=\nabla\times{\mathbf A}$ and a periodic potential 
$V({\mathbf r})$ 
\begin{equation}\label{ham}
  {\cal H} =
  {1\over{2m}}\left({\mathbf p}+{e\over c}{\mathbf A}\right)^2
     + V({\mathbf r})\,.
\end{equation}
 Brown and Zak's concepts can be generalized to a local gauge of the vector 
potential ${\mathbf A}$ \cite{flo97a,flo98}, $N$-dimensional lattices 
\cite{flo96}, and a spatially inhomogeneous magnetic field \cite{flo00}. This 
paper exploits the fact that after imposing the Born--von K\'arm\'an (BvK)
periodic 
conditions the magnetic translations form a finite-dimensional  projective 
representation of the 2D translation group. Kronecker products of irreducible 
projective representations can be applied to description of multi-particle 
states \cite{flo99,flo99a}.

Considering problems, which involve the magnetic field ${\mathbf H}$ determined
by the vector potential ${\mathbf A}$, one has to keep in mind that some
results may depend on a chosen gauge, though physical properties should
be gauge-independent. Two gauges are most frequently used in description of 
the 2D electron systems: the Landau gauge with ${\mathbf A}=[0,xH,0]$ and the
antisymmetric one with ${\mathbf A}=({\mathbf H}\times{\mathbf r})/2$. The 
relations between these gauges was discussed in the earlier article 
\cite{flo98}, so this problem is left out in the present considerations. 
However, 
it should be underlined that a form representation matrices depends on chosen 
gauge and, moreover, obtained representations are inequivalent, what means,
among others, that their bases are not related by a unitary transformation.
Since it is a symmetry adapted basis what results from presented material, 
then it is important to stress that the Landau gauge is assumed and obtained
results can not be immediately applied to other gauges.

\section{Irreducible projective representations of the 2D translation group}

All finite-dimensional irreducible projective representations of the 2D
translation for a given BvK period $N$ are labeled by numbers $n, l, k_1, k_2$,
where $n$ is a divisor of $N$, $0<l<n$ is mutually prime with $n$, and 
${\mathbf k}=(k_1,k_2)$ labels irreducible representations of 2D translation
group with the BvK period $N/n$, so $k_1,k_2=0,1,\dots,(N/n)-1$. In the special
case $n=1$ matrix elements are given by the following formula \cite{flo99a}
\begin{equation}\label{djk}
  D^l_{jk}[n_1,n_2]=\delta_{j,k-n_2}\omega_N^{ln_1j}\,,
\end{equation}
 where $l$ is mutually prime with $N$, $\omega_N=\exp(2\pi{\mathrm i}/N)$, 
$j,k,n_1,n_2=0,1,\dots,N-1$ (so
all expressions are calculated modulo $N$), 
and $[n_1,n_2]$ denotes a vector of the 2D 
translation group (strictly speaking, their coordinates in the lattice basis 
$\{{\mathbf a}_1,{\mathbf a}_2\}$). For the sake of simplicity the 
considerations are limited to this case and the presented results correspond
to the limit of high magnetic fields, {\it i.e.} there is no Landau level
mixing. 
The other special case $n=N$ leads to the standard (vector) representations 
of the translation group, {\it i.e.}
  \begin{equation}\label{krep} 
   D^{\mathbf k}[n_1,n_2]=\omega_N^{k_1n_1+k_2n_2}\,. 
  \end{equation}
 It has been shown in the earlier paper \cite{flo99a} that for a given magnetic
field ${\mathbf H}$ the indices $l$ and $n$ are related with the charge of a
moving particle. In the considered case it is assumed that the representation
$D^l$ corresponds to a hole, whereas $D^{-l}$ --- to an electron. The vector
representation $D^{\mathbf k}$ corresponds to a neutral particle or to an
electron-hole pair, what is a case considered here.
Hence, trions $X^\pm$ are related with a Kronecker product of three
representations: $D^{\pm l}\otimes D^{\pm l}\otimes D^{\mp l}$.

\section{Trion states and their symmetrization}\label{direct}

The trions $X^\pm$ are charged excitons with the charge $\pm e$ equal to that
of a single hole or electron, so, from the group-theoretical point of view,
their states have to transform as vectors of the irreducible projective 
representation $D^{\pm l}$. Since representations $D^{\pm l}$ are 
$N$-dimensional
then the following decomposition is true (see also \cite{flo99a})
 \begin{equation}\label{deco}
  D^{\pm l}\otimes D^{\pm l}\otimes D^{\mp l} = N^2 D^{\pm l}\,.
 \end{equation} 
 This relation expressed in terms of the basis vectors has a form
  \begin{equation}\label{basis}
   \ket{w}_{\pm}^{pq}=\sum_{stu}a_{stu,w}^{pq}\ket{stu}_{\pm\pm\mp}\,,
  \end{equation}
  where $s,t,u,w,p,q=0,1,\dots,N-1$, $\ket{w}_{\pm}$ is a state of the trion 
$X^\pm$, $\ket{stu}_{\pm\pm\mp}$ is a three-particle state with $s,t$ labeling 
states
of two holes (electrons), $u$ --- a label of a single electron (hole), and 
the pair $p,q$ plays a role of a repetition index. The state $\ket{w}_\pm$ 
should behave as a basis
vector of the representation $D^{\pm l}$, so
  \begin{equation}\label{bvec1}
    D^{\pm l}[n_1,n_2]\ket{w}_\pm
   = \omega_N^{\pm ln_1(w-n_2)} \ket{w-n_2}_\pm\,.
  \end{equation}
 It is satisfied if $a_{stu,w}^{pq}= \delta_{s,w+p}\delta_{t,w+q} 
\delta_{u,w+p+q}$, or
 \begin{equation}\label{wstate}
   \ket{w}_{\pm}^{pq}=\ket{(w+p)\,(w+q)\,(w+p+q)}_{\pm\pm\mp}\,.
 \end{equation}
 Namely, for each  pair $p,q$ the product (\ref{deco}) acts
 on a the ket $\ket{w}_\pm^{pq}$ as follows
 \begin{eqnarray*}
    D^{\pm l}\otimes D^{\pm l}\otimes D^{\mp l}[n_1,n_2]\ket{w}_\pm^{pq} &=&
    D^{\pm l}[n_1,n_2]\ket{w+p} 
    D^{\pm l}[n_1,n_2]\ket{w+q} 
    D^{\mp l}[n_1,n_2]\ket{w+p+q}  \\
   &=& \omega_N^{\pm ln_1(w-n_2)} 
  \ket{(w+p-n_2)(w+q-n_2)(w+p+q-n_2)}_{\pm\pm\mp} \\
  &=& \omega_N^{\pm ln_1(w-n_2)}\ket{w-n_2}_{\pm}\,.
 \end{eqnarray*}
  For $p=q$ the obtained states are symmetric with respect to the 
transposition of identical particles (holes or electrons). In the other
cases ($p\neq q$) it is easy to form symmetric and antisymmetric combinations
 \begin{equation}\label{sym}
   \ket{w}_{\pm}^{pq\pm}=2^{-1/2}(\ket{w}_{\pm}^{pq}\pm\ket{w}_{\pm}^{qp})\,,
 \end{equation}
 where now $q>p=0,1,\dots,N-1$. Completely antisymmetric
 states can be constructed by introducing antisymmetric and symmetric
 spin functions, respectively. Non-interacting trions, therefore, will be
describe by a similar Hamiltonian as for free holes (electrons) ({\it cf.} Eq.\ 
(\ref{ham}))  with a modified effective mass and, if necessary, 
an appropriate potential $V({\mathbf r})$. However, the degeneracy of energy
levels is $N^2$ times larger.

Another form of the irreducible basis can be obtained when one consider
at first conjugation of two representations and next conjugation of the
resultant representation with the third one. The first step can be done in 
two ways: (i) two identical representations are conjugated, {\it i.e.} one 
considers a product $D^l\otimes D^l$ or $D^{-l}\otimes D^{-l}$ or (ii) states
of a pair hole-electron are determined. The first method is more 
interesting since it can be used in problems where pairs of identical particles
come into play ({\it e.g.} superconducting states). However, the considerations
are a bit more complicated since the parity of $N$ has to be taken into 
account \cite{flo97}. Hence, in this paper the second possibility will be
investigated, whereas the first is left for the further works.

\subsection{States of a hole-electron pair}
To begin with a hole-electron pair (in a general case: a particle-antiparticle
pair) is taken into account. Since such a pair has the charge 
zero, then its behavior in an external magnetic field should be similar 
(up to effective mass etc.) to this of a non-charged
particle. It means that in the algebraic picture ordinary (vector)
representations of the translation group should appear. This is confirmed
by the brief outlook presented above: projective representations 
used to labeling of electron and hole states differ in the sign of $l$ only.
When both particles move in the same magnetic field then corresponding 
representations are $D^{+l}$ and $D^{-l}$ (of course, the BvK
period $N$ is identical in both cases). In this case 
matrix elements of these representations are given by (\ref{djk})
 and those of their product are
  \begin{eqnarray}
  \nonumber
     (D^{+l}\otimes D^{-l})_{jj',kk'}[n_1,n_2] 
   &=& D^{+l}_{jk}[n_1,n_2]D^{-l}_{j'k'}[n_1,n_2] 
   \\
   &=& \delta_{j,k-n_2}\delta_{j+k',k+j'} \omega_N^{ln_1(j-j')}\,.
  \label{prodele}
  \end{eqnarray}
 The product of representations $D^{+l}\otimes D^{-l}$ is a reducible 
representations
which decomposes into irreducible one-dimensional vector representations 
(\ref{krep})
  \begin{equation}\label{deco1}
D^{+l}\otimes D^{-l} = \bigoplus_{\mathbf k} D^{\mathbf k} \,.
  \end{equation} 
 There is no need to use a repetition index,because each representations
appears only ones. We are looking for such $N^2$ linear combinations
 $$
  \ket{1}_{\mathbf k}=\sum_{s,t}a_{st}^{\mathbf k}\ket{s}_+\ket{t}_-
 $$ 
 that each behaves as a basis vector of a given irreducible representations 
$D^{\mathbf k}$. It can be shown that $a_{st}^{\mathbf k}=\delta_{t,s-xk_1}
\omega_N^{sk_2}$, where $x$ is the inverse of $l$ modulo $N$ (since $l$ is 
mutually prime with $N$, then $x$ is well-determined), {\it i.e.} $xl=1\bmod N$. 
Namely one obtains 
  \begin{eqnarray*}
   (D^{+l}\otimes D^{-l})[n_1,n_2]
 \ket{1}_{\mathbf k}
&=& \sum_s \omega_N^{sk_2}\,
D^{+l}[n_1,n_2] \ket{s}_+ D^{-l}[n_1,n_2]\ket{s-xk_1}_-
\\
&=&\sum_s 
 \omega_N^{sk_2}
\omega_N^{ln_1(s-n_2)} \ket{s-n_2}_+
    \omega_N^{ln_1(n_2+xk_1-s)} \ket{s-xk_1-n_2}_-
\\
&=&
    \omega_N^{k_1n_1+k_2n_2}
\sum_{s'} 
 \omega_N^{s'k_2}
  \ket{s'}_+ \ket{s'-xk_1}_-
   = D^{\mathbf k}[n_1,n_2]\ket{1}_{\mathbf k}\,.
  \end{eqnarray*}

\subsection{Trion states}

The results presented in the previous section yield 
  \begin{equation}\label{onek}
\ket{1}_{\mathbf k}= \sum_s \omega_N^{sk_2} 
 \ket{s\,(s-xk_1)}_{+-}\,,
\qquad xl = 1 \bmod N\,.
  \end{equation}
  Therefore,
 $$
  D^{+l}\otimes D^{-l}\otimes D^{\pm l} = \bigoplus_{\mathbf k} D^{\mathbf k} 
\otimes D^{\pm l}
$$
and trion states $\ket{w}_{\pm}^{\mathbf k}$, $w=0,1,\dots,N-1$, can be 
written as
  $$
  \ket{w}_{\pm}^{\mathbf k}= \sum_{s,t,u} b^{w;{\mathbf k}}_{stu}
\ket{stu}_{+-\pm}\,,
  $$ 
 where $s,t$ label states of a pair and $u$ --- those of the second hole
(electron) for a trion $X^\pm$.
 The repetition index ${\mathbf k}$ follows from the way in which final 
states are constructed: at first states $\ket{s}_+$ and $\ket{t}_-$ are 
conjugated
to the states $\ket{1}_{\mathbf k}$ according with Eq.~(\ref{onek}) and
next linear combinations of pairs $\ket{1}_{\mathbf k}\ket{u}_\pm$ are 
considered. Therefore one can write
 $$
  \ket{w}_{\pm}^{\mathbf k}= \sum_u c^{w;{\mathbf k}}_u
\ket{1}_{\mathbf k}\ket{u}_\pm\,.
 $$
 Since for each ${\mathbf k}=(k_1,k_2)$, $k_1,k_2=0,1,\dots,N-1$ one has
\cite{flo99,flo99a}
  $$
     D^{\mathbf k}\otimes D^{\pm l}=D^{\pm l}\,,
  $$
 then each such product yields states $\ket{w}_{\pm}$, but the coefficients
$c_u^w$ depend on ${\mathbf k}$ and are given as
  $$
   c_u^{w;{\mathbf k}}=\omega_N^{-wk_2}\delta_{u,w\mp xk_1}\,,
  $$
  where, again, $xl=1\bmod N$, so
 \begin{equation}\label{wtrion} 
  \ket{w}_{\pm}^{\mathbf k}= 
\omega_N^{-wk_2}
\ket{1}_{\mathbf k}\ket{w\mp xk_1}_\pm\,.
  \end{equation}
 Taking into account equations (\ref{bvec1}) and (\ref{krep}) one obtains
  \begin{eqnarray*}
   (D^{\mathbf k}\otimes D^{\pm l})[n_1,n_2]
  \ket{w}_{\pm}^{\mathbf k} &=& \omega_N^{-wk_2}
   D^{\mathbf k} [n_1,n_2] \ket{1}_{\mathbf k}
D^{\pm l} [n_1,n_2] \ket{w\mp xk_1}_\pm \\
&=&  
   \omega_N^{-(w-n_2)k_2}
  \omega_N^{\pm ln_1(w-n_2)}
\ket{1}_{\mathbf k}
\ket{w\mp xk_1-n_2}_\pm \\
   &=& \omega_N^{\pm ln_1(w-n_2)} 
 \ket{w-n_2}_\pm
 =D^{\pm l}[n_1,n_2]\ket{w}_{\pm},.
\end{eqnarray*}
 Equations (\ref{onek}) and (\ref{wtrion}) lead to the final expression
($xl = 1 \bmod N$, $N^{-1/2}$ is a normalization factor)
 \begin{equation}\label{nnsw}
  \ket{w}_{\pm}^{\mathbf k}= 
N^{-1/2}\omega_N^{-wk_2}
 \sum_s \omega_N^{sk_2} \ket{s\,(s-xk_1)\,(w\mp xk_1)}_{+-\pm}\,,
 \end{equation}
 In such a state there is a kind of symmetry between an electron and a hole
forming the neutral pair electron-hole, but there is no symmetry between
two holes (electrons) in a trion $X^\pm$. Since there are $N^2$ trion states
labeled by $w$ then it is possible to construct states symmetric and 
antisymmetric with respect to particles transposition. One of possible ways
is presented in Sec.~\ref{direct} and it is easy to determined a 
transformation between the obtained bases. 

 Let us consider states of a trion $X^+$. The results of Sec.~\ref{direct}
read
 \begin{equation}\label{tplus}
   \ket{w}_+^{pq}=\ket{(w+p)\,(w+q)\,(w+p+q)}_{++-}\,,
 \end{equation}
 where the first two indices correspond to hole states and third to a state
of an electron. In the above presented formula (\ref{nnsw}) holes are
labeled by the first and the third indices, whereas the middle one corresponds
to an electron. Therefore, to calculate scalar products the order of indices
has to be changed in one of these formulae. Having this done one obtains
 \begin{eqnarray}
   {}_+^{\mathbf k}\scpr{w}
  {w}_+^{pq} 
&=& N^{-1/2}\omega_N^{wk_2} \sum_s \omega_N^{-sk_2} 
{}_{+-+}\scpr{s\, (s-xk_1)\,(w-xk_1)}{(w+p)\,(w+p+q)\,(w+q)}_{+-+}
\nonumber\\
&=& N^{-1/2}  \omega_N^{-pk_2} \delta_{q+xk_1,0}
 \label{trans}
 \end{eqnarray}

In the simplest case $N=2$ this formula reads (the unique representation is
obtained for $l=x=1$; ${\mathbf k}=(k_1,k_2)$)
  \begin{eqnarray*}
   \ket{w}_+^{00} &=& 2^{-1/2}\left(
      \ket{w}_+^{(0,0)} + \ket{w}_+^{(1,0)}
\right) \,,\\
   \ket{w}_+^{01} &=& 2^{-1/2}\left(
     \ket{w}_+^{(0,0)} - \ket{w}_+^{(1,0)}
\right)\,, \\
   \ket{w}_+^{10} &=& 2^{-1/2}\left(
   \ket{w}_+^{(0,1)} + \ket{w}_+^{(1,1)}
\right) \,,\\
   \ket{w}_+^{11} &=& 2^{-1/2}\left(
    \ket{w}_+^{(0,1)} - \ket{w}_+^{(1,1)}
\right)\,. 
  \end{eqnarray*}
 The second and the third formulae can be symmetrized what yields the following 
expressions 
  \begin{eqnarray*}
   \ket{w}_+^{01+} &=& 2^{-1}\left(
     \ket{w}_+^{(0,0)} - \ket{w}_+^{(1,0)}
  + \ket{w}_+^{(0,1)} + \ket{w}_+^{(1,1)}
\right)\,, \\
   \ket{w}_+^{01-} &=& 2^{-1}\left(
     \ket{w}_+^{(0,0)} - \ket{w}_+^{(1,0)}
  - \ket{w}_+^{(0,1)} - \ket{w}_+^{(1,1)}
\right)\,. 
  \end{eqnarray*}

\section{Final remarks}
  The presented considerations have shown that free trions should
behave in similar way way as free electrons or holes. However, due
to their internal structure the degeneracy is higher and there are
many possibilities to construct states $\ket{w}_\pm$, two of which
have been discussed above. In these simplified considerations there are
no interactions between trions or Landau level mixing and, moreover, the spin
or angular momentum numbers. Taking into account spins will allow to construct
states completely antisymmetric with respect to the permutational symmetry.
Such problem has been discussed lately by Dzyubenko {\it et al.} \cite{Dzy}
for the case of free trions ({\it i.e.} without a periodic potential, so
there is no discrete translational symmetry).
A sum of indices in the RHS of (\ref{tplus}), taking into account signs
of charges, is $(2w+p+q)-(w+p+q)=w$, what is equal to the index in the LHS 
of this equation. This is the same result as presented in \cite{Dzy}, where
the total angular momentum projection of a trion equals $(n_1-m_1)+(n_2-m_2)
-(n_3-m_3)$, where $(n_j-m_j)$ is the total angular momentum projection for
holes ($j=1,2$) and an electron ($j=3$), with $n$ and $m$ being the Landau 
level and the oscillator quantum numbers, respectively. It is interesting
that Dzyubenko {\it et al.} obtained their results in the antisymmetric
gauge ${\mathbf A}=({\mathbf H}\times{\mathbf r})/2$, whereas in the presented
considerations the Landau gauge has been used. It confirms that the physical
properties are gauge-independent. On the other hand, the actual form of wave 
functions is not discussed here, but the relations between
representations and their product are taken into account only. These relations
are independent of the matrix representations and, similarly, the form of
resultant basis is independent of the function form: for a given BvK period
$N$ and any gauge irreducible projective representations are $N$-dimensional
and their action on basis vectors are similar (up to a factor system) 
\cite{brow,zak,flo98,fish}.

% now the references. delete or change fake bibitem. delete next three
%   lines and directly read in your .bbl file if you use bibtex.

\end{document}